# Variation of the fine-structure constant: an update of statistical analyses with recent data.

L. Kraiselburd[1] ⋆, S. J. Landau[2]⋆⋆, and C. Simeone[2,3]⋆⋆⋆

[1] Grupo de Astrofísica, Relatividad y Cosmología, Facultad de Ciencias Astronómicas y Geofísicas, Universidad de La Plata, Paseo del Bosque S/N 1900 La Plata, Pcia de Buenos Aires, Argentina
[2] Instituto de Física de Buenos Aires, CONICET - Universidad de Buenos Aires, Ciudad Universitaria - Pab. 1, 1428 Buenos Aires, Argentina
[3] Departamento de Física, FCEN, UBA, Ciudad Universitaria - Pab. 1, 1428 Buenos Aires, Argentina
e-mail: `lkrai@fcaglp.unlp.edu.ar, slandau@df.uba.ar, csimeone@df.uba.ar`

**ABSTRACT**

*Aims.* We analyze the consistency of different astronomical data of the variation in the fine-structure constant obtained with KECK and VLT.
*Methods.* We tested the consistency using the Student test and confidence intervals. We splited the data sets in to smaller intervals and grouped them acording to redshift and angular position. Another statistical analysis is proposed that considers phenomenological models for the variation in $\alpha$.
*Results.* Results show consistency for the reduced intervals for each pair of data sets and suggests that the variation in $\alpha$ is significant at higher redshifts.
*Conclusions.* Even though the dipole model seems to be the most accurate phenomenological model, the statistical analyses indicate that the variation in $\alpha$ might be depending on both redshift and angular position.

## 1. Introduction

The time variation of fundamental constants has motivated much theoretical and experimental research since the large-number hypothesis (LNH) proposed by Dirac (1937). The high predictive power of the LNH induced many research papers and suggested new sources of variation. Among them, the attempt to unify all fundamental interactions resulted in the development of multi-dimensional theories, for instance string-motivated field theories (Wu and Wang, 1986; Maeda, 1988; Barr and Mohapatra, 1988; Damour and Polyakov, 1994; Damour et al., 2002a,b), related brane-world theories (Youm, 2001a,b; Palma et al., 2003; Brax et al., 2003), and (related or not) Kaluza-Klein theories (Kaluza, 1921; Klein, 1926; Weinberg, 1983; Gleiser and Taylor, 1985; Overduin and Wesson, 1997), which predict not only an energy dependence of the fundamental constants, but also a dependence of their low-energy limits on cosmological time. Many observational and experimental efforts have been made to establish constraints on these variations. The experimental research can be grouped into astronomical and local methods. These last include geophysical methods such as the natural nuclear reactor which operated about $1.8 \times 10^9$ years ago in Oklo, Gabon (Damour and Dyson, 1996; Petrov et al., 2006; Gould et al., 2006), the analysis of natural long-lived $\beta$ decayers in geological minerals and meteorites (Olive et al., 2004), and laboratory measurements such as comparisons of rates between clocks with different atomic numbers (Prestage et al., 1995; Sortais et al., 2001; Marion et al., 2003; Bize et al., 2003; Fischer et al., 2004; Peik et al., 2004). The astronomical methods are based mainly on the analysis of high-redshift quasar absorption systems. Most of the reported data are, as expected, consistent with null variation of fundamental constants. On the other hand, Dzuba et al. (1999) and Webb et al. (1999) proposed the many multiplet method (MMM), which, by comparing different transitions in the same absorption cloud, gains an order of magnitude in sensibility with respect to previously reported data. Using this method, Dzuba et al. (1999), Webb et al. (1999) and Murphy et al. (2003) have reported observations made with the Keck telescope that suggest a lower value of the fine structure constant ($\alpha$) at high redshift than its local value. However, an independent analysis performed with VLT/UVES data gave null results (Srianand et al., 2004; Chand et al., 2004). More recently, Murphy et al. (2007) discussed these latest results and claimed that they were biased toward zero variation and underestimated errors due to the fitting procedure. Srianand et al. (2007) replied to these comments, and reanalysed their observations. These authors concluded that the results presented earlier were robust but recognized that the corresponding errors were larger than first reported. Contrary to these results, a recent analysis using VLT/UVES data also suggests a variation in $\alpha$ but in the opposite sense, that is, $\alpha$ appears to be larger in the past (Webb et al., 2011). The discrepancy between Keck/HIRES and VLT/UVES is yet to be resolved. In particular, the two studies relied on data from different telescopes observing different hemispheres. In addition, King et al. (2012) suggested that the Keck/Hires and VLT/UVES observations can be made to be consistent when the fine structure constant is spatially varying. A similar analysis of the same data performed by Berengut et al. (2012) points in the same direction. In a previous paper (Landau and Simeone, 2008) we have pointed out that results calculated from the mean value over a wide redshift range (or cosmological time scale) are at variance with those obtained considering shorter intervals. In this paper, we re-analyze the available data obtained with the many multiplet method with the Keck and VLT telescopes, using the statistical tools and method introduced in the previous paper. We also take into account the suggestion made by King et al. (2012): we group the data by redshift and angular position and apply the statistical tools to intervals shorter than those considered before. Furthermore, we propose another statistical method for studying the discrepancy between Keck and VLT data. We consider three phenomenological models for the $\alpha$ variation: i) null variation, ii) time variation of $\alpha$ equal to the mean value of each data set, and iii) spatial variation of $\alpha$ following the dipole model proposed by King et al. (2012). In each case we compute the amount of data of each group that lie within the Gaussian distribution corresponding to each model. In section 2 we review the statistical tools introduced in our previous paper and present the new ones. In section 3 we report the results of applying the Student test to the available data and calculating confidence intervals in those cases where the number of available data are not enough to perform the Student test. We also report the amount of availabe data that lie within 3 and $6 - \sigma$ of the Gaussian distribution corresponding to each phenomenological model. In Section 4 we present our conclusions.

⋆ posdoctoral fellow of CONICET
⋆⋆ member of the Carrera del Investigador Científico y Tecnológico, CONICET
⋆⋆⋆ member of the Carrera del Investigador Científico y Tecnológico, CONICET

## 2. Statistical tools

The problems to be addressed are i) whether, for given redshift intervals and independently of their angular position, two compared experiments are consistent or not; ii) whether, for different angular positions but independently of the redshift, two experiments are consistent or not. The corresponding procedure accordingly is a test for the difference between two population means, which involves a statistic defined in terms of two sample means and two sample variances. Now, in some cases, one of the experiments includes for a given redshift interval in case i) or for a given angular position interval in case ii) very few data, so that one cannot reasonably define a sample mean and a sample variance. In this situation, the procedure to be followed instead involves a confidence interval constructed from the sample values of the experiment with a number of data that do make a statistical treatment possible. In what follows, we introduce these two approaches and discuss the choice of the sample size, which, consequently, determines the width of the redshift intervals in case i) and the width of angular intervals in case ii).

### 2.1. Student test

The consistency of two experiments is taken as the null hypothesis, which is formulated as

$$H_0 : \mu_1 - \mu_2 = 0, \qquad (1)$$

where $\mu_1$, $\mu_2$ are the (unknown) population means of each experiment for a definite redshift interval in case 1) and for a given angular interval in case 2). The sample sizes for the available observational data are not expected to be large, and the true variances are not known; therefore to test the hypothesis we need to work with the sample variances and use a $t$ test. When the number of values within each sample are not equal, the usual $t$ test is not robust to departures from normality or from equality of variances. We then adopt an approximate test based on the statistic (Devore, 1995)

$$T = \frac{\overline{X}_1 - \overline{X}_2}{\sqrt{\frac{S_1^2}{m} + \frac{S_2^2}{n}}}, \qquad (2)$$

where $\overline{X}_1$, $\overline{X}_2$ are the sample mean values for the given interval and $m, n$ are the data numbers of each sample. The weighted sample variances $S_1^2, S_2^2$ are given by (Kendall et al., 1994; Brandt, 1989)

$$S_j^2 = \sum_i p_i \left(x_i - \overline{X}\right)^2, \qquad (3)$$

$$p_i = \frac{\frac{1}{e_i^2}}{\sum \frac{1}{e_i^2}}, \qquad (4)$$

where $e_i$ is the observational error. The null hypothesis is rejected (so the two experiments compared are not considered consistent) when the statistic lies in the so-called rejection region ($RR$); for a two-tailed test, for which the alternative hypothesis is $\mu_1 - \mu_2 \neq 0$, the rejection region is defined by

$$RR : \begin{cases} T \leq -t_{\frac{\lambda}{2},\nu} \\ T \geq \phantom{-}t_{\frac{\lambda}{2},\nu}, \end{cases} \quad (5)$$

while for a one-tailed test we have

$$RR : T \leq -t_{\lambda,\nu}, \quad (6)$$

for the alternative hypothesis $\mu_1 - \mu_2 < 0$ and

$$RR : T \geq t_{\lambda,\nu}, \quad (7)$$

for the alternative hypothesis $\mu_1 - \mu_2 > 0$. In the expressions above the number of degrees of freedom $\nu$ is given by the rounded value of

$$\tilde{\nu} = \frac{\left(\dfrac{S_1^2}{m} + \dfrac{S_2^2}{n}\right)^2}{\dfrac{(S_1^2/m)^2}{m-1} + \dfrac{(S_2^2/n)^2}{n-1}}, \quad (8)$$

and $\lambda$ is the (approximate) level of the test (Brownlee, 1960). Thus, $\lambda$ is the approximate probability of a type I error, that is, the probability of rejecting the null hypothesis when it is true. In practice, we used an algorithm that yields a level $\lambda^*$ such that the obtained value of the statistic lies within the associated rejection region, which is given by substituting $\lambda^*$ in Eqs. (5), (6) and (7). Hence, at level $\lambda$ the null hypothesis should be rejected when $\lambda^* \leq \lambda$.

### 2.1.1. Choice of the intervals

One important point with our analysis is the selection of the intervals to be tested: the ideal situation would be to avoid any biasing associated with the arbitrary choice of location and size of the intervals. To achieve this independence, we have proposed the following procedure in our previous paper (Landau and Simeone, 2008): The first interval (to be considered to apply the test) starts at a redshift/angle $a$ with fixed $b$ width. The following $i$ intervals start at redshift/angles $a + i*c$ with the same width; $c = 0.1$ for the selection according to redshift, and $c = 0.025$ when selecting according to angular position. After testing all these intervals, we changed the value of $b$ and performed the same analysis again. In all cases we considered a minimum number of data in each interval as a necessary condition for applying the test. For the intervals that did not fulfill this condition on the number of data, we used another method, which we describe in section 2.2.

### 2.2. Confidence intervals

In some cases, one of the experiments to be tested included very few data and did not allow us to define a sample mean and a sample variance for a given redshift interval for case i) or an angular interval for case ii). In other cases, the amount of data is statistically too low to consider the results to be reliable. For these we introduced a different procedure: assuming that for a given interval a group of data 1 allows a statistical treatment, while a group of data 2 does not. To test the consistency of a given observation 2 against observation 1, we constructed an interval $I$ of confidence $100\,P\%$ from the values of group 1. Then, if the null hypothesis is true, $P = 1 - \lambda$ is the probability that the result of an observation of group 2 lies within this confidence interval, and the null hypothesis should be rejected at level $\lambda$ when this is not the case. The confidence interval was then centered at the mean value $\overline{X}$ of sample 1, and its width was determined by the complement of the rejection region of a two-tailed test. Thus, under the same hypothesis of the preceding subsection, we have

$$I = \left(\overline{X} - t_{\frac{\lambda}{2},\,n-1}\frac{S}{\sqrt{n}}\,;\,\overline{X} + t_{\frac{\lambda}{2},\,n-1}\frac{S}{\sqrt{n}}\right), \quad (9)$$

where $n$ is the number of values of sample 1, and $S^2$ is again the weighted sample variance. As before, the choice of the $t$ distribution is motivated by the size of the samples, which is not expected to be large. In practice, we chose a level $\lambda$ and the algorithm yielded a confidence interval for this level. Then, we compared the confidence interval obtained from group 1 with each single reported value for the variation of $\alpha$ obtained from group 2.

### 2.3. Type II error and sample size

A type II error is the failure to reject a false null hypothesis; its associated probability is usually denoted by $\beta$. While the probability $\lambda$ of type I error can be fixed independent of the population or sample values, the calculation of $\beta$ requires the choice of a definite alternative hypothesis; that is, to determine $\beta$, the inequality $\mu_1 \neq \mu_2$ must be specialized as a definite equality $\mu_1 - \mu_2 = \delta$. Under the assumption of an approximate normal distribution and a null hypothesis $\mu_1 - \mu_2 = 0$, for a two-tailed test and sample sizes $m$ and $n$, the probability $\beta$ is given by

$$\beta = \Phi\left(z_{\lambda/2} - \delta/S\right) - \Phi\left(-z_{\lambda/2} - \delta/S\right), \quad (10)$$

where $z_\lambda$, $z_\beta$ are obtained by inverting a normal N(0,1) distribution, $S^2 = S_1^2/m + S_2^2/n$, and $\Phi$ is the normal cumulative distribution function. For a one-tailed test, when the alternative hypothesis is $\mu_1 - \mu_2 > 0$, we have

$$\beta = \Phi\left(z_\lambda - \delta/S\right), \quad (11)$$

and for the alternative hypothesis $\mu_1 - \mu_2 < 0$ we have

$$\beta = 1 - \Phi\left(-z_\lambda - \delta/S\right). \quad (12)$$

We see that the probability $\beta$ measures for a given alternative hypothesis and certain sample sizes whether the data have led to a too conservative result or not. More precisely, a high value of $\beta$ implies that the data variance is larger than the difference between the null hypothesis and a given alternative hypothesis, which makes it difficult to resolve them. Small sample sizes will often lead to such high values of $\beta$. Conversely, we could choose a desired level $\beta$, and for a definite value of the alternative hypothesis then obtain an estimate of the required sample size $n$. For samples of similar sizes an approximate analytical expression is given in terms of the normal distribution; for a one-tailed test for the mean of a population (or for the associated confidence interval) we have (see Ref. (Devore, 1995))

$$n \simeq (z_\lambda + z_\beta)^2 \frac{S_1^2 + S_2^2}{\delta^2}. \qquad (13)$$

For a two-tailed test an analogous expresion applies, with $\lambda/2$ instead of $\lambda$. Of course, this would slightly underestimate the size $n$, because we should really use the $t$ distribution, which is less peaked than the normal distribution, but this is not relevant if one is not interested in an exact result. From a statistical point of view, a possible criterion to estimate the appropriate sample sizes is to limit the probability of type II error. In this approach, the choice of the approximate sample size would then be determined by the type II error that one is to admit for a given departure from the null hypothesis, this departure being measured by comparison with the sample variance.

### 2.4. Test of the null, mean value and dipole models

We also propose another way to analyze data on varying $\alpha$, which is similar to some of the analyses performed by King et al. (2012). However, in this case, we add the data reported by Srianand (2013, private comm.) (which come from the reanalysis (Srianand et al., 2007) of 21 observations made by Chand et al. (2004)) to the discussion and include the null hypothesis as a possible model. We assumed three phenomenological models for the $\alpha$ variation: i) null variation, ii) the value of $\alpha$ in the past was different from its actual value and is a fixed number, and iii) the variation of $\alpha$ follows the dipole model proposed by King et al. (2012). We proceeded as follows: for each phenomenological model we constructed the normal distribution asociated with its mean value and standard deviation. Then, we calculated the amount of data from each group that lies within the 3 and 6 standard deviations of the proposed normal distributions. We tested each data group separately. For the null distribution we took the standard deviation associated with the mean value reported by Srianand et al. (2007). For ii), we considered the mean value and standard deviation reported by each group. For the dipole model we have one distribution for each value of $\theta$; and the value of the dipole coordinates is equal to the one obtained by King et al. (2012) considering each group of data separately.

## 3. Results

In this section we analyze results of applying the Student test and/or confidence interval method described in Sect. 2 to recent astronomical data. Next we describe the data considered in this paper. Bounds on the variation in $\alpha$ were established from different methods (see Sect.1). However, the most stringent and abundant data are these obtained with the many multiplet method (Webb et al., 2001; Murphy et al., 2001). On the other hand, only two research groups (Webb et al. and Srianand et al.) have applied this method using two different telescopes: Keck and the Very Large Telescope (VLT). Therefore, we consider the following three groups of data to perform our statistical analysis:

**Group I**: Data obtained with the Keck telescope by Murphy et al. (2003) and King et al. (2012) (141 data points).
**Group II**: Data obtained with the VLT by King et al. (2012) (153 data points).
**Group III**: Data obtained with the VLT by Srianand (2013, private comm.). As discussed in Section 1, Srianand et al. (2007) have recognized that the data in Chand et al. (2004) should be reanalyzed. Therefore, we are considering for our analyses the mean value and enlarged errors provided by Srianand (2013, private comm.) and Srianand et al. (2007) (21 data points).

Furthermore, we recall that the statistical analyzes were performed considering data grouped by redshift and angular position.

### 3.1. Redshift

To apply the Student test grouping the data according to redshift, we considered datasets where the lowest total number of data is equal to 12 (n $\geq$ 12). The total redshift interval to be tested is $(0.440, 2.795)$; we applied the test to shorter intervals of equal width ($\Delta z = 0.35$) as described in section 2. Fig. 1 shows the value of $\lambda*$ obtained from the comparison of data from group I with data from group II. Not all values of $\lambda^*$ corresponding to the redshift interval $(1.470, 2.795)$ are higher than 0.025 and therefore this interval was discarded from the consistency interval. We also performed the same test and changed the value of the interval width $\Delta z$ and obtained similar results. Considering $\Delta z = 0.30$ we were able to test the interval $0.545 < z < 2.695$ (to fulfill the requirement $n \geq 12$ for both data sets); values of $\lambda^*$ are lower than 0.025 in the interval $(1.795, 2.695)$ while taking $\Delta z = 0.40$ we were able to test the redshift interval $0.345 < z < 2.845$ and obtained values of $\lambda^*$ lower than 0.025 in the redshift interval $(1.170, 2.845)$. These results give evidence that the variation in $\alpha$ may be relevant at higher redshifts, as was pointed out by Webb et al. (2003). On the other hand, the corresponding probability type II error ($\beta$) is shown in Fig. 2. The value of $\beta$ was obtained using the normal distribution (one-tail test, see section 2); we considered the alternative hypothesis equal to the mean value of $\frac{\Delta\alpha}{\alpha}$ obtained by each group: $\mu_1 = -0.6 \times 10^{-5}$ (Keck) and $\mu_2 = 0.2 \times 10^{-5}$ (VLT). In all cases the value of $\beta$ was higher than 0.1, showing the difficulty to distinguish the null hypothesis from the alternative hypothesis with the present data sets. We also performed the same analysis, but constrained the number of data to n ≥ 15 and n ≥ 18 to obtain get lower values of $\beta$; and changing the width interval as described above. However, the analyses showed that the value of $\beta$ did not change significantly, while increasing the lowest number of data in each data set involves ruling out many intervals from the possible intervals to be tested.

Since the requirements of applying the Student test are not always fulfilled by the available data, calculating of confidence intervals is a useful tool for testing consistency among data on varying $\alpha$. We chose $\lambda = 0.025$ and built a confidence interval for a group of data and compared the results with each single reported value of another author. Again, we decided on the bin size as the shortest interval centered at the reported value, which contains $n$ data. The criterion for analyzing the results was the same as in our previous paper (Landau and Simeone, 2008): *If all confidence intervals overlap with the reported value of the other author, we conclude that the whole interval is consistent, while if none of them overlap, we conclude that the interval is inconsistent. If there are some confidence intervals that do not overlap with the respective reported interval, the corresponding redshift interval is excluded from the consistency interval.* This is a conservative criterion, because we are probably overestimating the discarded intervals.

Table 1 shows the confidence intervals calculated for the data from group II for redshift intervals centered on each value of data from group III with the same telescope and containing at least 12 data points (we also calculated the confidence intervals for $n \geq 15, 18$ and discussed these results above). From Table 1 it follows that the confidence intervals calculated for $z = 1.348$, and $z = 2.022$ do not overlap with the corresponding reported intervals. Therefore, the redshift intervals $(1.278, 1.419)$ and $(1.935, 2.110)$ should be discarded from the consistency interval because the data from group II used to calculate the confidence intervals belong to this interval. Therefore, from the confidence intervals analysis we conclude that eight data points from group III are consistent with 53 data points from group II over the redshift interval $(0.142, 1.278)$ while seven data points from group III are consistent with 42 data points from group II over the redshift interval $(1.419, 1.935)$. Furthermore, four data points from group III and 18 data points from group II are consistent also along the redshift interval $(2.100, 2.429)$. The last column of Table 1 shows the type II error probability ($\beta$); in all cases this value is higher than 0.1, which indicates how difficulty it is to distinguish the null hypothesis from the alternative hypothesis with the present data sets. We also calculated the confidence intervals for $n \geq 15$ and $n \geq 18$ in an attempt to improve the value of $\beta$. However, the values of $\beta$ did not change significantly, while including more data in each confidence interval using the present data set leads to an enlarged redshift interval for which the confidence interval is calculated.

We also compared the data points obtained with the Keck Telescope by Murphy et al. (2003) and King et al. (2012) (group I) with those obtained by Srianand (2013, private comm.) with the VLT (group III) using confidence intervals. Table 2 shows that the redshift interval $(1.233, 1.838)$ should be discarded because the confidence intervals calculated for $z = 1.348$, $z = 1.555$ and $z = 1.657$ do not overlap with the corresponding reported intervals. Accordingly, the data points $z = 1.439$ and $z = 1.637$ from group III should be discarded as well. Consequently, the redshift intervals $(0.179, 1.233)$ and $(1.838, 2.457)$ are consistent, where there are 6 and 7 data points from group III, and 54 and 28 from group I respectively. A similar analysis was performed previously ((Landau and Simeone, 2008)), however, in the present work, we considered the enlarged errors reported by Srianand (2013, private comm.). The $\beta$ values are again high, noting that more data points are needed to reduce the type II error probability.

### 3.2. Spatial variation

Motivated by the analysis performed by King et al. (2012) with data from the KECK and VLT telescopes that suggested spatial variation, we performed another statistical analysis using the Student test and confidence intervals. This time, however, data were choosen according to their angular position instead of their redshift. The data set were selected by their value of $\cos\theta = \boldsymbol{X} \cdot \boldsymbol{D}$, where $\boldsymbol{X}$ is the quasar position and $\boldsymbol{D}$ is the dipole direction obtained by King et al. (2012) considering the Keck and VLT datasets. The angular distribution of available data is as follows: Keck data (group I) come from the region where $-1 \leq \cos\theta \leq 0.5$, while the VLT (group II and group III) reports data from the region where $-0.5 \leq \cos\theta \leq 0.9$. Accordingly, the Student test can be applied to reduced data sets from both telescopes. Furthermore, we have to reduce the lowest total value of each data set (to apply the test) to $n \geq 6$. Figure 3 shows the results of the Student test performed for reduced data set comparing group I with group II within the interval $-0.225 < \cos\theta < 0.225$. The width of the intervals for which the Student test was applied is $\cos\theta = 0.075$.

It follows from Fig 3 that not all values of $\lambda^*$ within the interval $(0.075, 0.150)$ are higher than $0.025$. Since data belonging to this interval were used to perform the Student test of other intervals, we analyzed these intervals again. For the interval $(0.05, 0.075)$ there is only one data set reported by Keck and four data sets from the VLT. Therefore we calculated a confidence interval with the VLT data and compared the result with the reported data from Keck, finding consistency within this interval. For the interval $(0.150, 0.225)$ the Student test can be applied because there are seven data sets from Keck and 12 data sets from VLT. We obtained $\lambda^* = 0.17$ for this interval.

To complete our analysis we performed the Student test by changing the value of the width interval for which the test is applied. Considering a width interval $\cos\theta = 0.070$, the test can be applied to the interval $-0.225 < \cos\theta < 0.220$ and the values of $\lambda^*$ are lower than 0.05 in the interval $(0.075, 0.145)$. Furthermore, for a width interval of $\cos\theta = 0.080$ the test can be applied to the interval $-0.225 < \cos\theta < 0.230$ and we find consistency between both data sets for the intervals $(-0.225, 0.050); (0.155, 0230)$.

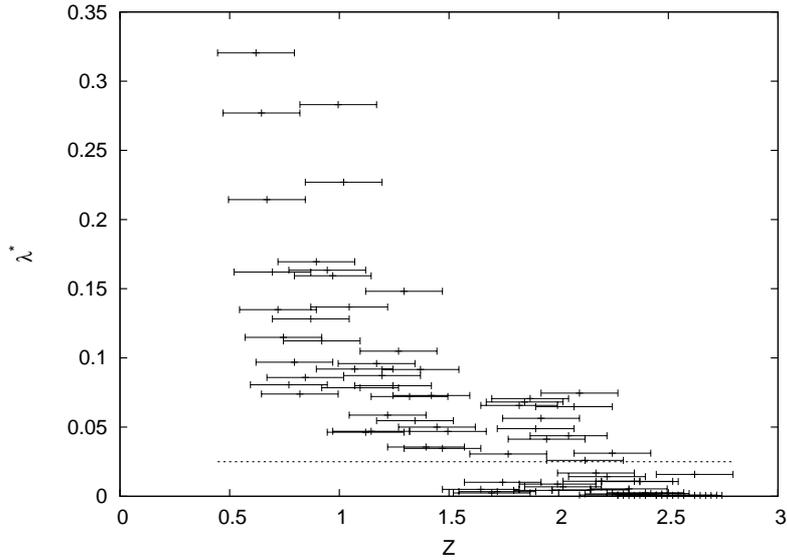

**Fig. 1.** Results of Student test comparing data obtained with the Keck telescope (Murphy et al., 2003; King et al., 2012)(group I) with data obtained with VLT (King et al., 2012) (Group II). Data sets are selected according to redshift; $\lambda*$ is the calculated level of the test for each redshift interval (the dotted line indicates the $\lambda* \leq 0.025$ rejection region).

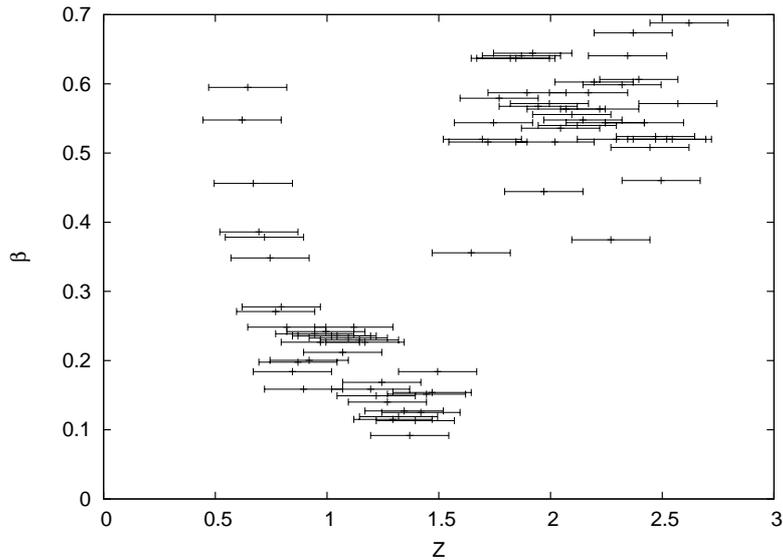

**Fig. 2.** Type II probability errors calculated for the Student test using $\lambda = 0.025$ for each redshift interval comparing results obtained with the Keck Telescope (Murphy et al., 2003; King et al., 2012) (group I) with results obtained with VLT (King et al., 2012) (group II). Data sets are selected according to redshift.

Figure 4 shows the type II error probability for the width interval 0.075; it follows that almost all values of $\beta$ are higher than 0.1, showing the need for more data to reduce this value. We also calculated the values of $\beta$ for other width intervals and found no significant difference with results shown in Fig. 4 .

As described in section 3.1, the requirements for applying the Student test are not always fulfilled by the available data, and therefore calculating confidence intervals is a useful tool for testing the consistency among data on varying $\alpha$. Again $\lambda = 0.025$ and we built a confidence interval for a group of data and compared the results with

**Table 1.** Confidence intervals for different redshifts for $n \geq 12$ comparing data obtained with the VLT by Srianand (2013, private comm.) (group III) with data obtained by King et al. (2012) with the same telescope (group II).

| $z$ | $\Delta z$ | Reported Interval | Confidence Interval | $\beta$ for $\alpha = 0.05$ |
|---|---|---|---|---|
| (1) | (2) | (3) | (4) | (5) |
| 0.452 | 0.621 | $(-0.92, -7.99 \times 10^{-2})$ | $(-1.42, 0.32)$ | 0.86 |
| 0.822 | 0.155 | $(-0.60, 1.20)$ | $(-0.41, 1.16)$ | 0.84 |
| 0.859 | 0.179 | $(-0.34, 1.04)$ | $(0.23, 1.64)$ | 0.82 |
| 0.873 | 0.164 | $(-5.00 \times 10^{-2}, 1.05)$ | $(0.29, 1.71)$ | 0.82 |
| 0.942 | 0.152 | $(-2.05, 0.35)$ | $(-0.82, 0.50)$ | 0.81 |
| 1.182 | 0.106 | $(-0.56, 0.96)$ | $(-0.92, 0.84)$ | 0.86 |
| 1.243 | 0.121 | $(-3.30, -0.50)$ | $(-0.67, 0.97)$ | 0.85 |
| 1.277 | 0.146 | $(-0.90, 0.90)$ | $(-0.88, 1.02)$ | 0.87 |
| 1.348 | 0.141 | $(1.25, 3.15)$ | $(-0.92, 0.25)$ | 0.78 |
| 1.439 | 0.178 | $(-1.01, 0.61)$ | $(-0.97, 0.15)$ | 0.77 |
| 1.555 | 0.147 | $(-0.31, 0.71)$ | $(-0.41, 1.90)$ | 0.88 |
| 1.636 | 0.114 | $(-0.90, 0.30)$ | $(-0.28, 2.10)$ | 0.89 |
| 1.637 | 0.116 | $(-0.40, 1.80)$ | $(-0.28, 2.10)$ | 0.89 |
| 1.657 | 0.141 | $(0.10, 1.10)$ | $(-0.11, 1.90)$ | 0.87 |
| 1.858 | 0.163 | $(-0.37, 0.77)$ | $(-0.61, 0.90)$ | 0.83 |
| 1.915 | 0.208 | $(-9.99 \times 10^{-2}, 1.30)$ | $(0.17, 1.23)$ | 0.75 |
| 2.022 | 0.175 | $(-2.30, -0.70)$ | $(8.13 \times 10^{-2}, 1.67)$ | 0.84 |
| 2.168 | 0.171 | $(-9.99 \times 10^{-2}, 1.50)$ | $(-0.13, 1.07)$ | 0.79 |
| 2.185 | 0.151 | $(5.00 \times 10^{-2}, 3.15)$ | $(-0.13, 1.07)$ | 0.79 |
| 2.187 | 0.155 | $(-0.70, 0.70)$ | $(-0.13, 1.07)$ | 0.79 |
| 2.300 | 0.257 | $(-0.70, 0.90)$ | $(-0.45, 1.01)$ | 0.83 |

Notes.- Columns: (1) redshift reported by Srianand (2013, private comm.) (group III);
(2) length of the redshift interval for which the confidence interval is calculated;
(3) single value reported by Srianand (2013, private comm.) in units of $10^{-5}$;
(4) calculated confidence interval from a group of data reported
by King et al. (2012)(group II) in units of $10^{-5}$; (5) type II error probability.

each single reported value of another author. The procedure for determining consistency intervals is the same as in the previous section.

For the interval where $\cos\theta > 0.225$, nine data sets are data reported by Keck and 90 reported by VLT. Table 3 shows the confidence intervals calculated for the data from group II for intervals centered on each value of data from group I, where its width is 0.1 and containing $n \geq 11$ data points. It should be noted that some of the Keck data belong to the same quasar, which reduces the calculation of confidence intervals of Table 3 to only five. In this way, there is consistency for the interval $0.225 < \cos\theta < 0.497$.

For the interval where $\cos\theta < -0.225$ there are 73 data sets from group I and 7 data sets from group II. Table 4 shows the confidence intervals calculated for the data from group I for intervals centered on each value of data from group II. The results show that the interval $(-0.428, -0.328)$ should be discarded from the consistency interval.

In summary, we analyzed the consistency over the interval $-0.454 < \cos\theta < 0.497$, comparing 87 data sets from group I with 121 data sets from group II. From the Student test and confidence interval calculation it follows that 68 data sets from group I (79% of the analyzed data) are consistent with 112 data sets from group II (93% of the analyzed data) over the intervals $-0.454 < \cos\theta < -0.429$, $-0.327 < \cos\theta < 0.05$, and $0.150 < \cos\theta < 0.497$.

Table 5 shows the confidence intervals calculated for the data from group II for $\cos\theta$ intervals (0.1 width) centered on each value of the data from group III with the same telescope and containing at least seven data points. The quantity of testable intervals is reduced to ten, because this is the number of quasars from which the Srianand (2013, private comm.) and Srianand et al. (2007) data came from. At first, it seems that the quasars with $\cos\theta = 0.372$ and $\cos\theta = 0.497$ are inconsistent with the calculated confidence intervals from group II. Therefore, the intervals $(0.322, 0.4215)$ and $(0.447, 0.547)$ should be

**Table 2.** Confidence intervals for different redshifts for $n \geq 12$ comparing data obtained with the VLT by Srianand (2013, private comm.) (group III) with data obtained with Keck by Murphy et al. (2003) and King et al. (2012) (group I).

| $z$ | $\Delta z$ | Reported Interval | Confidence Interval | $\beta$ for $\alpha = 0.05$ |
|---|---|---|---|---|
| (1) | (2) | (3) | (4) | (5) |
| 0.452 | 0.546 | $(-0.92, -7.99 \times 10^{-2})$ | $(-1.56, 0.99)$ | 0.89 |
| 0.822 | 0.153 | $(-0.60, 1.20)$ | $(-0.79, 0.75)$ | 0.84 |
| 0.859 | 0.137 | $(-0.34, 1.04)$ | $(-0.46, 0.40)$ | 0.68 |
| 0.873 | 0.109 | $(-5.00 \times 10^{-2}, 1.05)$ | $(-0.52, 0.32)$ | 0.67 |
| 0.942 | 0.150 | $(-2.05, 0.35)$ | $(-1.29, 6.08 \times 10^{-2})$ | 0.81 |
| 1.182 | 0.141 | $(-0.56, 0.96)$ | $(-1.48, 0.23)$ | 0.85 |
| 1.243 | 0.153 | $(-3.30, -0.50)$ | $(-1.73, -0.31)$ | 0.82 |
| 1.277 | 0.130 | $(-0.90, 0.90)$ | $(-1.21, -2.13 \times 10^{-2})$ | 0.79 |
| 1.348 | 0.231 | $(1.25, 3.15)$ | $(-1.17, -0.19)$ | 0.73 |
| 1.439 | 0.290 | $(-1.01, 0.61)$ | $(-1.27, -0.35)$ | 0.70 |
| 1.555 | 0.424 | $(-0.31, 0.71)$ | $(-1.28, -0.38)$ | 0.70 |
| 1.636 | 0.333 | $(-0.90, 0.30)$ | $(-1.60, -0.35)$ | 0.80 |
| 1.637 | 0.331 | $(-0.40, 1.80)$ | $(-1.60, -0.35)$ | 0.80 |
| 1.657 | 0.362 | $(0.10, 1.10)$ | $(-1.60, -0.35)$ | 0.80 |
| 1.858 | 0.264 | $(-0.37, 0.77)$ | $(-1.74, 0.53)$ | 0.88 |
| 1.915 | 0.294 | $(-9.99 \times 10^{-2}, 1.30)$ | $(-1.77, 0.64)$ | 0.89 |
| 2.022 | 0.237 | $(-2.30, -0.70)$ | $(-0.97, 1.13)$ | 0.87 |
| 2.168 | 0.283 | $(-9.99 \times 10^{-2}, 1.50)$ | $(-2.02, 0.36)$ | 0.89 |
| 2.185 | 0.249 | $(5.00 \times 10^{-2}, 3.15)$ | $(-2.02, 0.36)$ | 0.89 |
| 2.187 | 0.245 | $(-0.70, 0.70)$ | $(-2.02, 0.36)$ | 0.89 |
| 2.300 | 0.313 | $(-0.70, 0.90)$ | $(-2.35, -4.75 \times 10^{-2})$ | 0.88 |

Notes.- Columns: (1) redshift reported by (group III); (2) length of the redshift interval for which the confidence interval is calculated; (3) single value reported by group III in units of $10^{-5}$; (4) calculated confidence interval from a group of data reported by group I in units of $10^{-5}$; (5) type II error probability.

discarded. If we were too conservative, we should also reject the quasars with $\cos \theta = 0.330, 0.368, 0.519$ and the consistent intervals should be $(4.70 \times 10^{-4}, 0.322)$, $(0.422, 0.447)$ and $(0.547, 0.617)$ (63 data points from group II and 10 from group III). But if the Student test is aplied to the discarded intervals, these regions are consistent. That is, for the interval $(0.322, 0.4215)$ there are eight data points from Srianand (2013, private comm.) and 18 from King et al. (2012), being $\lambda* = 0.33$; and for $(0.447, 0.547)$ there are three data points from group III and 10 from group II, being $\lambda* = 0.22$. Therefore, we conclude that in this case the comparison between a single quasar data set and a confidence interval is not the appropiate tool for testing the consistency between groups of data. Thus, we performed a Student test using the same procedure as we applied for the comparison between Keck and VLT data reported by the group of Murphy et al. (see Figs 1 and 3) reducing the lowest total value of each data set to $n \geq 3$. Even though it is not ideal to perform a Student test with such a reduced data set, we considered that it is a better tool for the analysis than the confidence interval comparison presented above. Figure 5 shows the results of the Student test performed for the reduced data set comparing group III with group II within the interval $0.000 < \cos \theta < 0.625$. The width of the intervals for which the Student test was applied is $\Delta \cos \theta = 0.075$. All intervals derived from the Srianand (2013, private comm.) data are consistent with those built from the King et al. (2012) data. From Figure 6 it is clear that all $\beta$ values obtained in this case are again very high.

Table 6 shows the confidence intervals calculated for the data from group I for $\cos \theta$ intervals (0.1 width) centered on each value of the data from group III. Only ten data points from group III ($\sim 45\%$ of the data) can be used, and these in turn are able to form four intervals because some of them arise from the same quasar (29 data points from Keck can be compared (21% of the data)). The results show that the analyzed interval $(4.70 \times 10^{-4}, 0.380)$ is consistent. It can be seen in Tables 5 and 6 that the $\beta$

**Table 3.** Confidence intervals for data obtained with the VLT (King et al., 2012) (group II). Comparison with single data obtained with the Keck telescope (Murphy et al., 2003) and (King et al., 2012) (group I). Data sets are selected according to angular position.

| $\cos\theta$ | $n$ | Reported Interval | Confidence Interval | $\beta$ for $\alpha = 0.05$ |
|---|---|---|---|---|
| (1) | (2) | (3) | (4) | (5) |
| 0.307 | 30 | $(-2.19, 9.64 \times 10^{-2})$ | $(-7.10 \times 10^{-2}, 0.78)$ | 0.87 |
| 0.304 | 30 | $(-5.19, 1.40)$ | $(-7.10 \times 10^{-2}, 0.78)$ | 0.87 |
| 0.447 | 11 | $(-0.77, 1.33)$ | $(-0.41, 1.02)$ | 0.79 |
| 0.248 | 22 | $(-2.79, 5.53)$ | $(1.70 \times 10^{-2}, 1.15)$ | 0.91 |

Notes.- Columns: (1) $\cos\theta$ of data from the Keck telescope (group I); (2) number of data reported by the VLT inside the interval; (3) single value reported by Keck in units of $10^{-5}$; (4) calculated confidence interval from the VLT data in units of $10^{-5}$; (5) type II error probability.

**Table 4.** Confidence intervals for data obtained with the Keck telescope (Murphy et al., 2003) and (King et al., 2012) (group I). Comparison with single data obtained with the VLT (King et al., 2012) (group II). Data sets are selected according to angular position.

| $\cos\theta$ | $n$ | Reported Interval | Confidence Interval | $\beta$ for $\alpha = 0.05$ |
|---|---|---|---|---|
| (1) | (2) | (3) | (4) | (5) |
| $-0.249$ | 9 | $(-4.44, 0.58)$ | $(-2.25, -0.15)$ | 0.81 |
| $-0.306$ | 2 | $(-0.27, 4.64)$ | $(-12.20, 9.16)$ | 0.15 |
| $-0.378$ | 9 | $(2.31, 14.15)$ | $(-3.18, 1.08)$ | 0.15 |
| $-0.404$ | 11 | $(-7.49 \times 10^{-2}, 0.45)$ | $(-2.37, 1.05)$ | 0.82 |

Notes.- Columns: (1) $\cos\theta$ from data obtained with the VLT ; (2) number of data reported Keck Telescope inside the interval; (3) single value reported by the VLT in units of $10^{-5}$; (4) calculated confidence interval from the Keck data in units of $10^{-5}$; (5) type II error probability.

**Table 5.** Confidence intervals for data obtained with the VLT by King et al. (2012). Comparison with single quasar obtained with the VLT by Srianand (2013, private comm.). Data sets are selected according to angular position.

| $\cos\theta$ | $n$ | Reported Interval | Confidence Interval | $\beta$ for $\alpha = 0.05$ |
|---|---|---|---|---|
| (1) | (2) | (3) | (4) | (5) |
| 0.248 | 26 | $(-0.37, 0.77)$ | $(-4.55 \times 10^{-2}, 1.02)$ | 0.91 |
| 0.519 | 17 | $(9.99 \times 10^{-2}, 0.70)$ | $(-0.26, 1.21)$ | 0.75 |
| 0.368 | 18 | $(-0.54, 0.61)$ | $(-0.47, 0.27)$ | 0.74 |
| 0.330 | 30 | $(-1.94, 0.46)$ | $(-0.23, 0.42)$ | 0.77 |
| 0.552 | 9 | $(-0.49, 0.56)$ | $(-0.55, 1.69)$ | 0.82 |
| 0.567 | 11 | $(-1.01, 0.61)$ | $(-0.75, 1.25)$ | 0.79 |
| 0.266 | 31 | $(-0.68, 0.15)$ | $(1.18 \times 10^{-2}, 0.98)$ | 0.90 |
| $5.047 \times 10^{-2}$ | 7 | $(-5.87 \times 10^{-2}, 0.62)$ | $(-0.85, 1.21)$ | 0.91 |
| 0.497 | 10 | $(-2.30, -0.70)$ | $(-0.59, 1.52)$ | 0.84 |
| 0.372 | 18 | $(0.28, 0.59)$ | $(-0.47, 0.27)$ | 0.74 |

Notes.- Columns: (1) $\cos\theta$ from data obtained by Srianand (2013, private comm.); (2) number of data reported by King et al. (2012) inside the interval; (3) single value reported by Srianand (2013, private comm.) in units of $10^{-5}$; (4) calculated confidence interval from a group of data reported by King et al. (2012) in units of $10^{-5}$; (5) type II error probability.

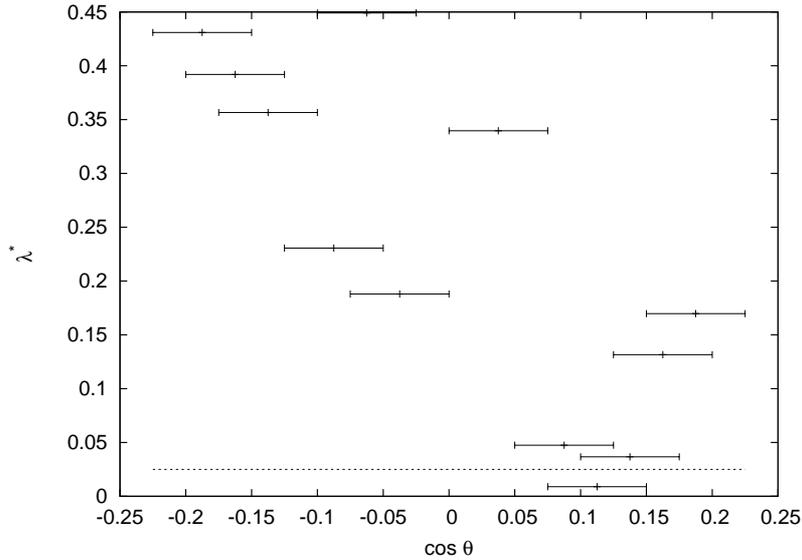

**Fig. 3.** Results of Student test comparing results obtained with the Keck telescope (Murphy et al., 2003) and (King et al., 2012) (group I) with results obtained with the VLT (King et al., 2012) (group II). Data sets are selected according to angular position, $\lambda*$ is the calculated level of the test for each interval (the dotted line indicates the $\lambda* \leq 0.025$ rejection region).

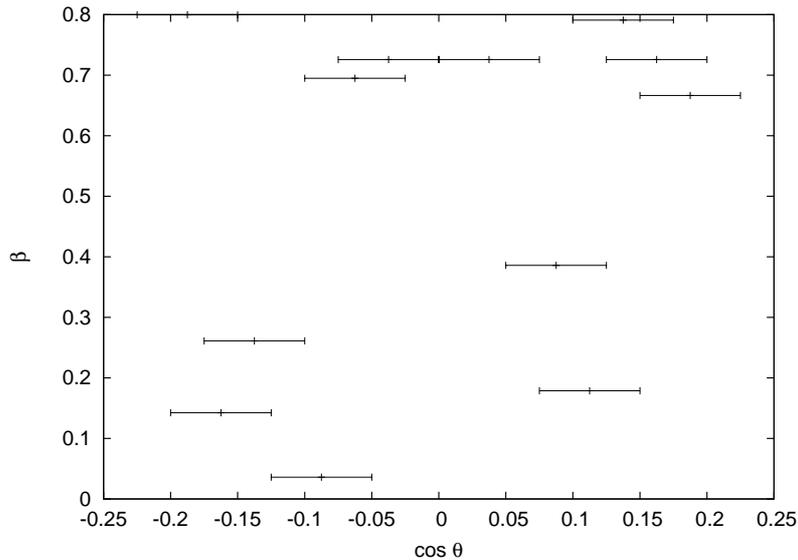

**Fig. 4.** Type II probability errors for each angular position interval calculated for the Student test using $\lambda = 0.025$ comparing results obtained with the Keck telescope (Murphy et al., 2003) and (King et al., 2012) (group I) with results obtained with the VLT (King et al., 2012) (group II). Data sets are selected according to angular position.

values are very high, suggesting again that a greater quantity of data is required to improve these values.

### 3.3. Phenomenological models

In section 2.4 we have described another method for analyzing the different groups of data on varying $\alpha$. The results are listed in Table 7. Although it can be noted that the three groups suit the dipolar model better than the other two models, there is still a large amount of data from group I and group II that is left out of both distributions. Furthermore, it should be noted data from the VLT (group II and group III) favor the dipole model over the other proposed phenomenological models, while data from the Keck telescope cannot distiguish between the dipole model and the null distribution.

### 4. Summary and Conclusions

We have performed different statistical analyses to test recent astronomical data that indicate a possible variation in the fine structure constant. We used statistical methods explored in a previous paper (Landau and Simeone, 2008), which involve the Student test and confidence intervals. This time, however, more data were added and grouped according to redshift and angular position. We also proposed some phenomenological models for the variation in $\alpha$ and computed the amount of data that lie within 3 and $6 - \sigma$ of the asociated Gaussian distribution.

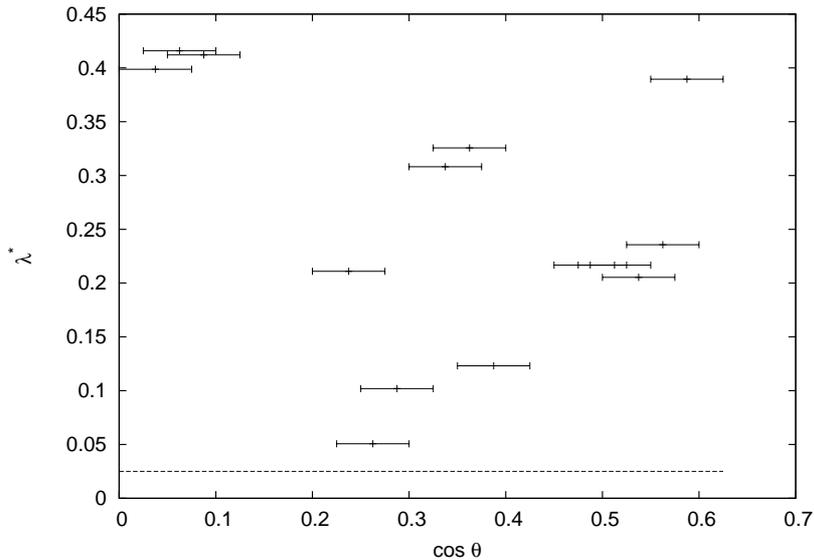

**Fig. 5.** Results of Student test comparing results obtained with the VLT by Srianand (2013, private comm.) (group III) with results obtained with the same telescope by (King et al., 2012) (group II). Data sets are selected according to angular position; $\lambda*$ is the calculated level of the test for each interval (the dotted line indicates the $\lambda* \leq 0.025$ rejection region).

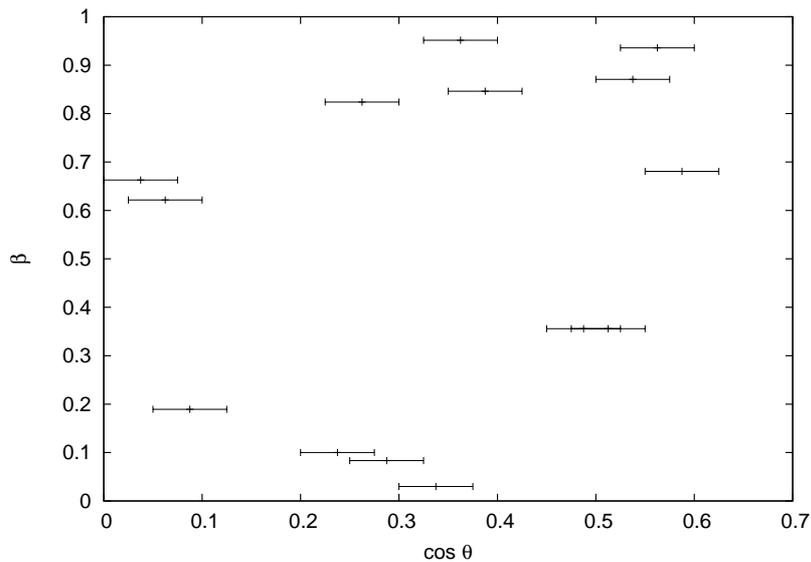

**Fig. 6.** Type II probability errors calculated for the Student test using $\lambda = 0.025$ for each angular position interval comparing results obtained with the VLT by Srianand (2013, private comm.) (group III) with results obtained with the same telescope by (King et al., 2012) (group II). Data sets are selected according to angular position.

While grouping the data according to redshift, results of the statistical analyses show that the variation in $\alpha$ is more relevant at higher redshift. From the analysis performed grouping the data according to angular position, results show consistency over most of the analyzed intervals. In all cases the value ofthe type II error ($\beta$) shows the need for more data to arrive powerful conclusions. Finally, the analysis of Gaussian distributions of the proposed phenomenological models suggests that although one cannot rule out a possible variation in $\alpha$, this may be due not only to the angular position but also to redshift. More data are required on this aspect as well.

**Table 6.** Confidence intervals for data obtained with the Keck telescope Murphy et al. (2003) and (King et al., 2012). Comparison with single quasar obtained with the VLT by Srianand (2013, private comm.). Data sets are selected according to angular position.

| $\cos\theta$ | $n$ | Reported Interval | Confidence Interval | $\beta$ for $\alpha = 0.05$ |
| (1) | (2) | (3) | (4) | (5) |
|---|---|---|---|---|
| 0.248 | 2 | $(-0.37, 0.77)$ | $(-27.83, 25.55)$ | 0.95 |
| 0.330 | 5 | $(-1.94, 0.46)$ | $(-3.97, 1.15)$ | 0.93 |
| 0.266 | 7 | $(-0.68, 0.15)$ | $(-3.28, 0.49)$ | 0.93 |
| $5.047 \times 10^{-2}$ | 13 | $(-5.87 \times 10^{-2}, 0.62)$ | $(-1.21, 6.61 \times 10^{-3})$ | 0.89 |

Notes.- Columns: (1) $\cos\theta$ from data obtained from the VLT; (2) number of data reported by Keck telescope inside the interval; (3) single value reported from the VLT in units of $10^{-5}$; (4) calculated confidence interval from Keck data in units of $10^{-5}$; (5) type II error probability.

**Table 7.** Percentage of data within the SD of the mean. The dipole hypothesis implies a spatial variation in $\alpha$ following the dipole model proposed by King et al. (2012), the coordinates of the dipole are those obtained by King et al. (2012) considering each group of data separately; the null hypothesis consists of a null mean and a standard deviation given by Srianand et al. (2007) and Srianand (2013, private comm.); and the mean value hypothesis contains the mean of the data group and the corresponding standard deviation.

| Data Group | $3\sigma$ | | | $6\sigma$ | | |
| | Dipole hyp | Null hyp | Mean value hyp | Dipole hyp | Null hyp | Mean value hyp |
|---|---|---|---|---|---|---|
| Group I | 21% | 21% | 15% | 45% | 38% | 30% |
| Group II | 46% | 7% | 16% | 71% | 37% | 30% |
| Group III | 95% | 33% | 38% | 95% | 81% | 81% |

## Acknowledgements

Support for this work was provided by PIP 0152/10 CONICET. The authors are grateful to Raghunathan Srianand for providing enlarged errors of the VLT data.